\documentclass[referee]{raa}
\usepackage{graphicx,times}
\usepackage{natbib}
\usepackage{amssymb,amsmath}
\bibpunct{(}{)}{;}{a}{}{,}

\newcommand{\nature}{\hbox{Nature}}
\usepackage[a4paper=true,dvipdfm=true,pagebackref=true]{hyperref}
\hypersetup{pdftitle = The title of my PDF, pdfauthor = My name, pdfsubject= The subject, pdfkeywords = keyword1 keyword2 keyword3}
\hypersetup{colorlinks = true, linkcolor = green, anchorcolor = red, citecolor = blue, filecolor = red, pagecolor = red, urlcolor = red}

\usepackage{color}

\begin{document}

   \title{Morphology and structure of BzK-selected galaxies at $z\sim2$ in the CANDELS-COSMOS field
$^*$
\footnotetext{\small $*$ Supported by the National Natural Science Foundation of China.}
}

 \volnopage{ {\bf 2014} Vol.\ {\bf X} No. {\bf XX}, 000--000}
   \setcounter{page}{1}

   \author{Guan-Wen Fang\inst{1,2}, Zhong-Yang Ma\inst{2,3}, Yang Chen\inst{2,3,4}, Xu Kong
      \inst{2,3}   }
   \institute{Institute for Astronomy and History of Science and Technology, Dali University, Dali 671003, China; {\it wen@mail.ustc.edu.cn}\\
	\and
Key Laboratory for Research in Galaxies and Cosmology, The University of Science
and Technology of China, Chinese Academy of Sciences, Hefei, Anhui, 230026, China\\
\and
Center for Astrophysics, University of Science and Technology of China, Hefei 230026, China; {\it xkong@ustc.edu.cn}\\
\and
SISSA, via Bonomea 265, I-34136 Trieste, Italy\\
\vs \no
   {\small Received 2014 ? ?; accepted 2014 ? ?}
}

\abstract{Utilizing a $BzK$-selected technique, we obtain 14550 star-forming galaxies (sBzKs)
and 1763 passive galaxies (pBzKs) at $z\sim2$ from the $K$-selected ($K_{\rm AB}<22.5$) catalog
in the COSMOS/UltraVISTA field. The differential number counts of sBzKs and pBzKs are consistent with
the results from the literature. Compared to the observed results,
semi-analytic models of galaxy formation and evolution provide too few (many)
galaxies at high (low)-mass end. Moreover, we find that the star formation rate (SFR) and stellar
mass of sBzKs follow the relation of main sequence.
Based on the HST/Wide Field Camera 3 (WFC3) F160W imaging, we find a wide range of
morphological diversities for sBzKs, from diffuse to early-type spiral structures, with
relatively high $M_{\rm 20}$, large size and low $G$, while pBzKs are elliptical-like compact
morphologies with lower $M_{\rm 20}$, smaller size and higher $G$, indicating the more
concentrated and symmetric spatial extent of stellar population distribution
in pBzKs than sBzKs. Furthermore, the sizes of pBzKs (sBzKs)
at $z\sim2$ are on average two to three (one to two)
times smaller than those of local early-type (late-type) galaxies with similar stellar mass.
Our findings imply that the two classes have different evolution modes and mass
assembly histories.
\keywords{galaxies: evolution --- galaxies: fundamental parameters --- galaxies: structure --- galaxies: high-redshift
}
}

   \authorrunning{G.-W. Fang et al. }            
   \titlerunning{Morphology and structure of BzK-selected galaxies}  
   \maketitle

%

\section{Introduction}

The formation and evolution of massive galaxies ($M_{\ast}>10^{10}~\rm M_{\odot}$) at $z\sim2$
is a hot issue of observational astronomy. There are many reasons, for instance the population of
Hubble sequence galaxies is already in place at $z\sim$ 1.5--2 (Fang et al. 2012), the universe star
formation rate density (SFRD) peaks at $z\sim2$ (Oesch et al. 2012), the specific SFR (sSFR)
evolves weakly at $z> 2$ (Gonz\'{a}lez et al. 2014), the galaxy's mass grows quickly at $1<z<3$
(Ilbert et al. 2013), luminous infrared galaxies ($L_{\rm 8-1000~\mu m}>10^{11}{~\rm L_{\odot}}$)
are more common at redshift $z\sim$ 1--3 (Murphy et al. 2013), and the number density of quasi-stellar
objects (QSOs) has a peak at $z\sim2$ (Richards et al. 2006).

Within the past decade, many novel techniques have been applied to select a sample of massive
galaxies at the epoch of $z\sim2$ and important investigation has been made in our understanding
of high-redshift galaxies (Chapman et al. 2003; Franx et al. 2003; Daddi et al. 2004; Kong et al. 2006;
Dey et al. 2008; Huang et al. 2009; Fang et al. 2012; Wang et al. 2012; Fang et al. 2014).
Such as submillimeter galaxies (SMGs with $F({\rm 850~\mu m})>0.5{~\rm mJy}$),
distant red galaxies (DRGs with $(J-K)_{\rm Vega}>2.3$),
ultraluminous infrared galaxies (ULIRGs; $L_{\rm 8-1000~\mu m}>10^{12}{~\rm L_{\odot}}$),
dusty-obscured galaxies (DOGs with $(R-[24])_{\rm Vega}>24$), etc.
Based on a simple two-color ($B-z$ and $z-K$) approach, Daddi et al. (2004) introduced the
criteria of $(z-K)_{\rm AB}=2.5$ and $BzK=(z-K)_{\rm AB}-(B-z)_{\rm AB}=-0.2$ to select
the sample of $z\sim2$ massive galaxies. Objects with $BzK>-0.2$ are classified as
star-forming galaxies (sBzKs). Sources with $BzK<-0.2$ and $(z-K)_{\rm AB}>2.5$ are
defined as passive galaxies (pBzKs).

Following the $BzK$ technique in Daddi et al. (2004),
 many groups selected large samples of BzKs (include sBzKs and
pBzKs) from different surveys (Kong et al. 2006; Lane et al. 2007; Blanc et al. 2008;
Hartley et al. 2008; Hayashi et al. 2009; Cassata et al. 2010; McCracken et al. 2010;
Onodera et al. 2010; Fang et al. 2012; Ryan et al. 2012). Furthermore, they also investigated
the physical properties of these galaxies, e.g., surface density, stellar mass, SFR,
near-infrared (NIR) spectroscopy, morphology, clustering, and size. So far the largest
sample of BzKs is from McCracken et al. (2010), they found that the clustering of BzKs
is much stronger than that of full $K$-limited ($K_{\rm AB}<22$) samples of field galaxies.
Moreover, the comoving correlation length of pBzKs ($r_{0}\sim 7~{\rm h^{-1} Mpc}$) is
larger than that of sBzKs ($r_{0}\sim 5~{\rm h^{-1} Mpc}$).

For galaxies at $z\sim2$, Hubble Space Telescope (HST)/Wide Field Camera 3 (WFC3)
NIR imaging can provide crucial clues to the rest-frame optical morphologies.
At such redshift HST/WFC3 NIR bands move beyond the Balmer
break (${\lambda}_{\rm rest}\geqslant 4000~\rm {\AA}$) to the redder wavelengths
and thus probe the light from the dominant stellar population of galaxy. This will
enable us to study the rest-frame optical morphologies and structures of BzKs at $z\sim2$.
By using HST/WFC3 NIR images in the Hubble Ultra Deep Field (HUDF), Cassata et al. (2010)
reported the structural features of 6 pBzKs, these galaxies in appearance have a relatively
regular morphology and smaller size than local ellipticals of analogous stellar masses.
The similar results were also found in Ryan et al. (2012), their pBzHs sample includes
30 passive galaxies at $z\sim2$ (use the $H$-band filter replaces the $K$-band filter).
Fang et al. (2012) measured nonparametric morphological parameters of BzKs (50 pBzKs and 173 sBzKs)
at $z\sim2$ in the Extended Groth Strip (EGS) field, and found that BzKs have both early and late types.

This paper will utilize HST/WFC3 F160W images to investigate the structural properties of
BzKs. Compared with previous works, we present the larger $BzK$ sample with high resolution
NIR imaging, and for the first time we study the size evolution of sBzKs.
We introduce the multi-band observations and data reduction of the COSMOS field in Section 2.
Section 3 describes the selection, redshift distribution and number counts of BzKs.
We show the $SFR-M_{\ast}$ correlation of BzKs in Section 4. We present the structural and
morphological properties of BzKs in Section 5 and 6, and summarize our results in Section 7.
Throughout this paper, we adopt a standard
cosmology $H_{\rm 0}=70$~km s$^{-1}$ Mpc$^{-1}$, $\Omega_\Lambda =0.7$, and $\Omega_{\rm M} = 0.3$.
All magnitudes use the AB system unless otherwise noted.

\section{Observations and data}

The Cosmic Evolution Survey (COSMOS) is targeted on a special area of the sky,
that has been observed with some of the world's most powerful telescopes on the
ground and in space, in wavelength range from X-rays through ultraviolet
and visible light, down to infrared and radio waves (Scoville et al. 2007).
In addition, it also includes optical/infrared spectroscopy survey using the
Keck DEIMOS and LRIS, Magellan IMACS, and VLT VIMOS spectrographs. More details of the
observation and data reduction in the COSMOS field can be found in
McCracken et al. (2012) and Muzzin et al. (2013). The multi-band photometry
data we use in our work is from the $K$-limited ($K<23.4$) catalog of the COSMOS/UltraVISTA field
provided by Muzzin et al. (2013), which is produced based on the NIR data from the
UltraVISTA DR1 (McCracken et al. 2012). Moreover, the derived physical parameters
we use in our study also come from the catalog of Muzzin et al. (2013), such as
stellar mass ($M_{\ast}$), SFR (${\rm SFR_{UV,uncorr}+SFR_{IR}}$), and photometric
redshift ($z_{\rm p}$, if there is no spectroscopic redshift available).

In this paper, we will utilize the latest released data of HST/WFC3
F160W high-resolution images in the CANDELS\footnote{Cosmic Assembly Near-IR
Deep Extragalactic Legacy Survey (CANDELS; Grogin et al. 2011 and
Koekemoer et al. 2011)}-COSMOS field to investigate the morphological features of BzKs
in our sample. The CANDELS/wide
COSMOS survey covers a total of $\sim$210~\rm arcmin$^2$ at $J-$
and $H-$band. The $5~\sigma$ limiting magnitude is 26.9 in the F160W filter.
HST/WFC3 F160W images were drizzled to $0''.06~{\rm pixel}^{-1}$.
Further details are in Grogin et al. (2011) for the survey and observational design,
and Koekemoer et al. (2011) for the data products.

\section{Selection, redshift distribution and number counts of BzKs}

To construct a sample of galaxies at $z\sim2$, we use the $BzK$ color criteria from
Daddi et al. (2004). The optical $B$- and $z$-band data is taken with  Subaru/SuprimeCam
($B_{j}$, $z^{+}$), while the $K$ data is from the VISTA/VIRCAM. In addition, we also apply
the color correction used by McCracken et al. (2010) to $B_{j}-z^{+}$ to keep unanimous with
the $BzK$ selection technique. As shown in Figure 1, blue dots represent 14550 sBzKs with
$BzK>-0.2$ (solid line) and $K<22.5$, and red dots correspond to 1763 pBzKs with $BzK<-0.2$ and
$z-K>2.5$ (dot-dashed line). Objects with $z-K<0.3(B-z)-0.5$ (dashed line), are classified
as stars.

\begin{figure*}
\centering
\includegraphics[angle=0,width=0.8\textwidth]{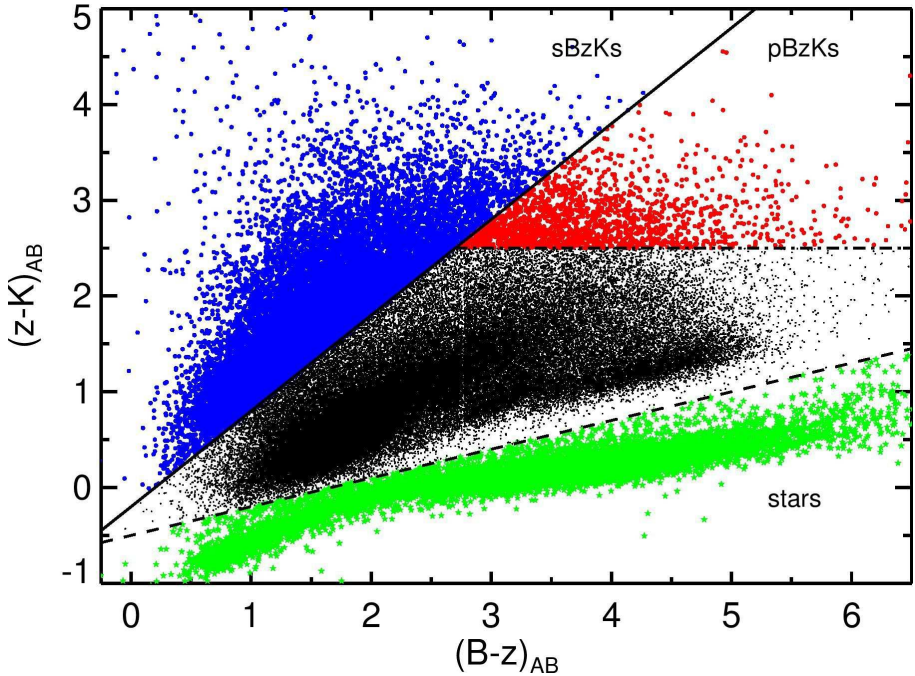}
\caption{$BzK$ two-color diagram for all objects in the COSMOS field.
Galaxies with $BzK>-0.2$ (solid line) and $K<22.5$ are selected as sBzKs.
Sources with $BzK<-0.2$ and $z-K>2.5$ (dot-dashed line) are defined as pBzKs.
The color criterion of star and galaxy separation is $z-K=0.3(B-z)-0.5$ (dashed line).
} \label{fig:bzks}
\end{figure*}

Figure 2 shows the redshift distribution for BzKs in the COSMOS field. For a sample of
galaxies with $K<22.5$, in Figure 2(a), we find that the $BzK$ color technique successfully selects
more than 80\% of galaxies at redshift $1.5<z<2.7$ (this fraction goes to more than 90\% at $1.6<z<2.6$),
indicating the $BzK$ criteria is quite effective in selecting galaxies at $z\sim2$.
For the massive galaxies ($M_{\ast}>10^{10}{\rm M_{\odot}}$) at $1.6<z<2.6$,
the percentage of objects selected as BzKs is $>90\%$ (the dotted red lines in Figure 2(a)).
The mean redshifts of sBzKs (Figure 2(b)) and pBzKs (Figure 2(c))
are $1.75\pm 0.48$ and $1.69\pm0.33$, respectively.
In the meantime, we use the extrapolation of the red-blue separation method
of Bell et al. (2004) to separate the red sequence and blue cloud in the rest-frame
$U-V$ vs. $M_{\rm V}$ diagram. About 91\% of pBzKs can be
roughly divided into red sequence, and most of sBzKs (84\%) are located in the blue
cloud region.

\begin{figure*}
\centering
\includegraphics[angle=0,width=0.6\textwidth]{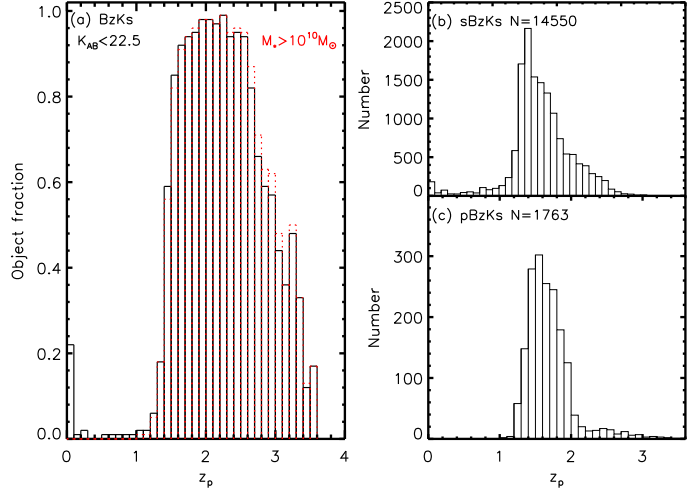}
\caption{(a) Fraction of BzKs (include sBzKs and pBzKs) in the total galaxy
sample of $K<22.5$. The dotted red lines show the fraction of massive
galaxies ($M_{\ast}>10^{10}{\rm M_{\odot}}$) selected as BzKs.
(b) Redshift distribution for sBzKs in our sample.
(c) Redshift distribution for pBzKs in our sample.
} \label{fig:zhist}
\end{figure*}

We calculated differential $K$-band number counts for all sBzKs and pBzKs in our sample
(see Figure 3). For comparison with the results from previous works, we also plotted
their data in Figure 3 (Deep3a-F and Daddi-F of Kong et al. 2006; Lane et al. 2007;
Blanc et al. 2008; Hartley et al. 2008; McCracken et al. 2010; Fang et al. 2012).
The dot-dashed lines represent the counts of quiescent galaxies (QGs) and star-forming galaxies
(SFGs) from the semi-analytic model (Kitzbichler \& White 2007). In general, our counts
agree with the results from the literature. Owing to the existence of photometric
offsets and cosmic variance, there is a discrepancy among different works for sBzKs and pBzKs counts.
Combining with the data from the literature, we confirm that the number counts of
pBzKs has a break at $K\sim 21.0$, a possible explanation for this lies in the small
redshift range of pBzKs (Kong et al. 2006). The redshift distribution of our
pBzKs sample also supports the finding, compared to sBzKs.
Compared with the observed results, the semi-analytic model predicts too few (many)
galaxies at high (low)-mass end.

\begin{figure*}
\centering
\includegraphics[angle=0,width=0.6\textwidth]{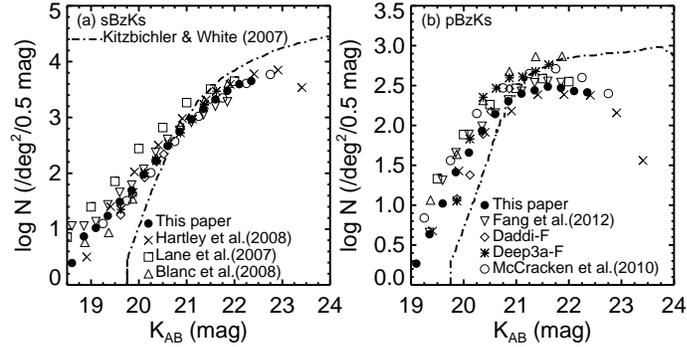}
\caption{Differential number counts of sBzKs and pBzKs in the COSMOS/UltraVISTA field.
The results from the literature and the model are also shown in this diagram.
Deep3a-F and Daddi-F from Kong et al. (2006).
} \label{fig:num}
\end{figure*}

\section{The stellar mass$-SFR$ correlation of BzKs}

For local star-forming galaxies, Brinchmann et al. (2004) found that there is a tight correlation
between $M_{\ast}$ and SFR (SFR${\propto}M^{\alpha}_{\ast}$), and called it main sequence (MS).
At redshift $0.5<z<3$, the MS is also confirmed (Daddi et al. 2007; Elbaz et al. 2007;
Rodighiero et al. 2011; Fang et al. 2012), but the slope ($\alpha$) ranges from 0.6 to 1.0
(relying on different sample and the approaches for calculating $M_{\ast}$ and SFR).
In Figure 4, we show the relation of $M_{\ast}$ and SFR of BzKs in the COSMOS field.
For sBzKs, a best-fit slope $\alpha=0.67\pm0.06$ (blue line) is found, in agreement with
those provided by Daddi et al. (2007) ($\alpha \sim 0.9$, gray line) and Rodighiero et al. (2011)
($\alpha \sim 0.79$, cyan line). The discrepancy for different slopes is due to the different
methods in deriving SFR (Rodighiero et al. 2014). On the other hand, we find that the SFRs and
stellar masses of pBzKs also show a correlation, but with lower SFRs compared to sBzKs
for a given stellar mass. Gray squares in Figure 4 represent sBzKs in Pannella et al. (2009).
From Karim et al. (2011), SFGs (gray triangles) with different mass and redshift bins
are also plotted in this figure.

\begin{figure*}
\centering
\includegraphics[angle=0,width=0.6\textwidth]{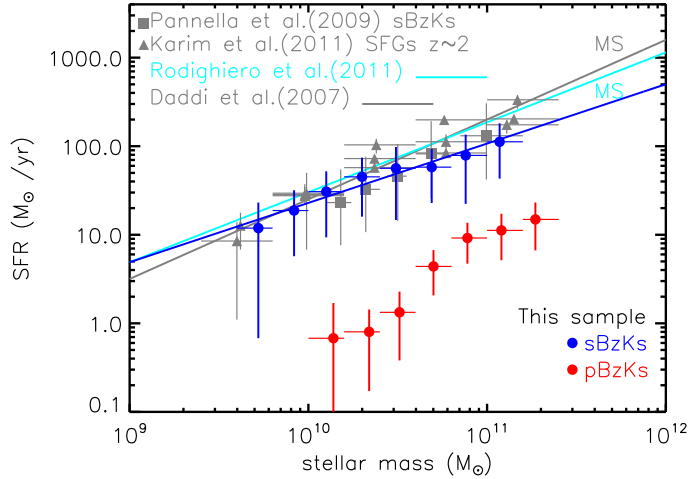}
\caption{Relationship of stellar mass vs. SFR for BzKs in the COSMOS field.
Solid squares and triangles represent the star-forming galaxies from
Pannella et al. (2009) and  Karim et al. (2011), respectively.
Gray and cyan lines correspond to the MS from Daddi et al. (2007)
and Rodighiero et al. (2011), respectively.
} \label{fig:sfrm}
\end{figure*}

\section{Structures of BzKs}

In order to analyze the structural properties of BzKs in the CANDELS-COSMOS field,
we employ the latest catalog\footnote{http://www.mpia-hd.mpg.de/homes/vdwel/candels.html}
(version 1.0) provided by van der Wel et al. (2012).
Galaxy sizes ($r_{\rm e}$) are measured from the HST/WFC3 F160W imaging.
Within a matched radius of $0''.5$ , we obtain the structural
parameters of 52 pBzKs and 378 sBzKs from van der Wel et al. (2012). The $M_{\ast}-r_{\rm e}$
relations are shown in Figure 5 of pBzKs and sBzKs, respectively. Shen et al. (2003)
provided the relations with $1\sigma$ dispersion for local late- and early-type galaxies
(LTGs and ETGs) as plotted in this figure. And for comparison, other data from the literature
are also shown in Figure 5(b) (Cassata et al.2010; Gobat et al. 2012; Ryan et al. 2012;
Szomoru et al. 2012). From Figure 5, we find that the sizes of pBzKs and massive sBzKs
at $z\sim2$ are smaller than their local counterparts at a fixed stellar mass.
Moreover, we also see a diversity of structural properties among BzKs, some sources are
similar to local galaxies, but there is also the existence of massive compact BzKs,
compared to present-day counterparts. Generally, pBzKs have elliptical-like compact structures
with low $r_{\rm e}$, while sBzKs are relatively extend and irregular, with higher $r_{\rm e}$.

\begin{figure*}
\centering
\includegraphics[angle=0,width=0.8\textwidth]{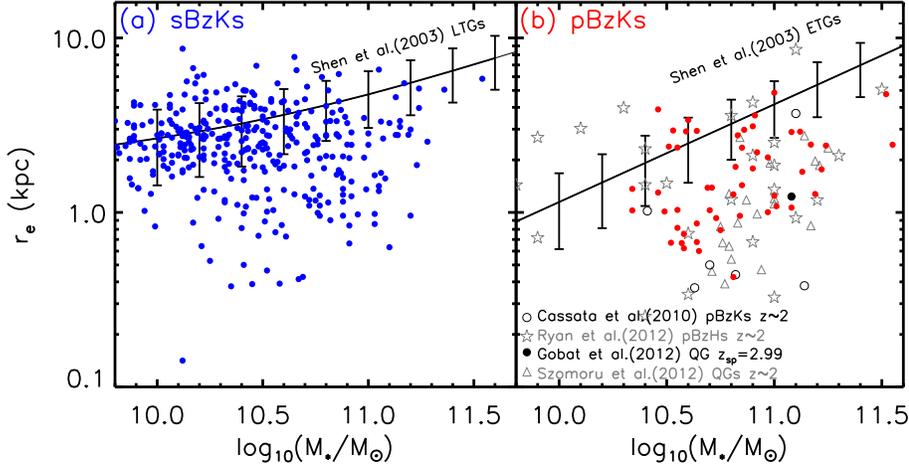}
\caption{Distribution of $M_{\ast}$ vs. $r_{\rm e}$ for BzKs ((a) sBzKs and (b) pBzKs)
in the CANDELS-COSMOS field. The results from the literature are also shown in
Figure 5(b) (Cassata et al.2010; Gobat et al. 2012; Ryan et al. 2012; Szomoru et al. 2012).
} \label{fig:rem}
\end{figure*}

To further investigate the size evolution with redshift for our BzKs sample at $z\sim 2$.
We show the sizes for pBzKs ($1.85\pm1.09$ kpc) and sBzKs ($2.63\pm1.36$ kpc)
in Figure 6, respectively. The effective radii of QGs and SFGs from the literature
are also plotted in this figure (Shen et al. 2003; Cassata et al.2010; Gobat et al. 2012;
Ryan et al. 2012; Szomoru et al. 2012; Fan et al. 2013; Patel et al. 2013;
Morishita et al. 2014). As shown in Figure 6, we find that the sizes of pBzKs
are a factor of $\sim 2-3$ smaller than those of local counterparts. For sBzKs,
the mean size is one to two times smaller than those of typical local LTGs with comparable mass.
Combined with the data points of Patel et al. (2013) and Morishita et al. (2014),
the difference of sizes for pBzKs and sBzKs indicates that the two classes
have different evolution processes and assembly histories, such as minor mergers
with a low increase in galaxy's mass and secular evolution without mergers
(or monolithic collapse mode). For the size evolution of pBzKs, our observations
support that the predictions from minor mergers. As regards the size growth of sBzKs,
a possible explanation lies in internal evolution through some quenched mechanisms
(AGN feedback or stellar winds).

\begin{figure*}
\centering
\includegraphics[angle=0,width=0.8\textwidth]{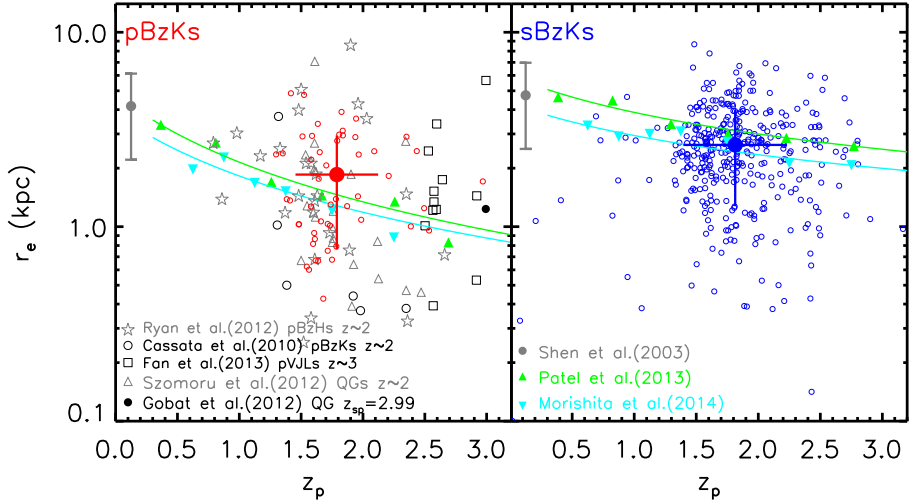}
\caption{Evolution of the effective radius ($r_{\rm e}$) with redshift
for pBzKs and sBzKs in our sample.
The effective radii of QGs and SFGs from the literature are also plotted in this figure.
The mean size of pBzKs is $1.85\pm1.09$ kpc (red solid circle), while sBzKs is
$2.63\pm1.36$ kpc (blue solid circle). $Left$: Green and cyan lines
correspond to $r_{\rm e}\propto(1+z)^{-1.16}$ (Patel et al. 2013)
and $r_{\rm e}\propto(1+z)^{-1.06}$ (Morishita et al. 2014), respectively.
$Right$: Green and cyan lines represent $r_{\rm e}\propto(1+z)^{-0.63}$ (Patel et al. 2013)
and $r_{\rm e}\propto(1+z)^{-0.56}$ (Morishita et al. 2014), respectively.
} \label{fig:rez}
\end{figure*}

\section{Morphologies of BzKs}

Using the WFC3 on board the HST, CANDELS provided
high resolution F160W imaging data ($0''.06~{\rm pixel}^{-1}$). In our work, we utilize
HST/WFC3 F160W images to study the morphological diversities of $z\sim2$ BzKs.
To clearly analyze their structural features, we calculated nonparametric morphological
parameters of galaxy, such as Gini coefficient ($G$; the relative distribution of the galaxy
pixel flux values) and $M_{\rm 20}$ (the second-order moment of the brightest 20$\%$ of the
galaxy's flux) (Abraham et al. 1996; Lotz et al. 2004).

In Figure 7, the red and blue circles represent pBzKs and sBzKs, respectively. Meanwhile, based on
the values of $G$ and $M_{\rm 20}$ of 52 pBzKs and 378 sBzKs, their stamp images ($3''\times3''$)
are also plotted in this figure. As shown in Figure 7, we find that pBzKs in appearance have
regular and compact (like spheroid), and that show low $M_{\rm 20}$ and high $G$ in rest-frame optical morphology.
For sBzKs, there is a wide range of morphological diversities, including clumpy, irregular, extended, and
early-type spiral-like morphologies, but most of them show diffuse structures,
with high $M_{\rm 20}$ and low $G$. That indicates the more concentrated and symmetric spatial extent of
 stellar population distribution in pBzKs than sBzKs. Furthermore, we derived the mean values of $G$ and $M_{\rm 20}$
for pBzKs and sBzKs, corresponding to ($0.63, -1.70$) and ($0.51, -1.49$), respectively.
Our findings further imply that passive galaxies and star-forming galaxies have
different evolution modes and mass assembly histories.

\begin{figure*}
\centering
\includegraphics[angle=0,width=0.7\textwidth]{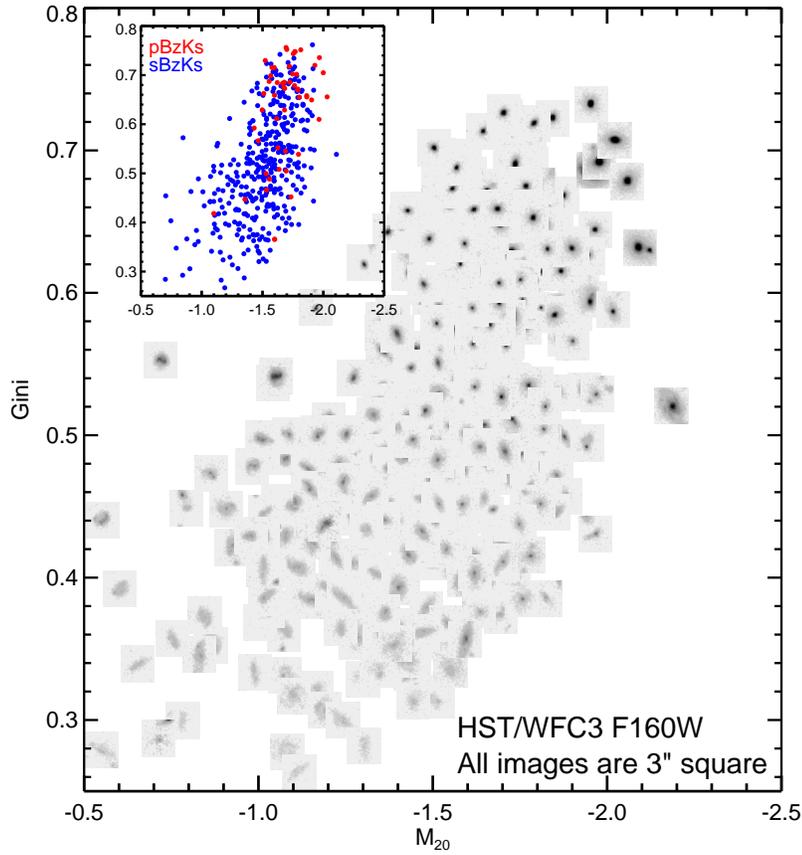}
\caption{Distribution of BzKs in the $M_{\rm 20}$ vs. Gini coefficient plane.
The red and blue circles represent pBzKs and sBzKs, respectively.
} \label{fig:gm20}
\end{figure*}

\section{Summary}

Based on  a $BzK$-selected technique, we present 14550 star-forming galaxies (sBzKs)
and 1763 passive galaxies (pBzKs) at $z\sim2$ from the $K$-selected ($K_{\rm AB}<22.5$) catalog
of the COSMOS/UltraVISTA field. Utilizing data from HST WFC3/F160W imaging in the CANDELS-EGS field,
we investigate the morphological and structural diversities of these galaxies. Our findings are
as follows:

(1) We find that the $BzK$ color technique successfully selects
$>$ 80\% galaxies at redshift $1.4<z<2.7$ (this fraction $>$ 90\% at $1.6<z<2.6$),
indicating the $BzK$ criteria is a quite effective galaxy selection method at $z\sim2$.
Moreover, for the massive galaxies ($M_{\ast}>10^{10}{\rm M_{\odot}}$) at $1.6<z<2.6$,
the percentage of sources selected as BzKs is $>90\%$.

(2) The differential number counts of sBzKs and pBzKs agree with
the results from the literature. Compared with the observed results,
models for galaxy formation and evolution provide too few (many) galaxies at high (low)-mass end.

(3) We find that the star formation rate (SFR) and stellar mass of sBzKs follow the
relation of main sequence (SFR${\propto}M^{0.67\pm0.06}_{\ast}$).
pBzks also shows correlation between SFRs and stellar masses but with too low SFRs for pBzKs.
Moreover, about 91\% of pBzks can be roughly divided into red sequence.

(4) We find that the sizes of pBzKs ($1.85\pm1.09$ kpc) and sBzKs ($2.63\pm1.36$ kpc)
at $z\sim2$ are smaller than their local counterparts at a fixed stellar mass.
Moreover, we also see a diversity of structural properties among BzKs, some sources are
similar to local galaxies, but there is also the existence of massive compact BzKs,
compared to present-day counterparts. The sizes of sBzKs are larger than
pBzKs in general, even in high-mass systems, but some have very compact structures,
with $r_{\rm e}<1 {\rm kpc}$.

(5) We find a wide range of morphological diversities for sBzKs, from extended or
diffuse to early-type spiral-like structures, while pBzKs are relatively regular and
compact (like spheroid). Moreover, we calculate the mean values of $G$ and $M_{\rm 20}$
for pBzKs and sBzKs, corresponding to ($0.63, -1.70$) and ($0.51, -1.49$), respectively.
The sBzKs show high $M_{\rm 20}$ and low $G$, which indicates less
concentrated and symmetric spatial distribution of the stellar mass of sBzKs
at $z\sim2$, comparing to pBzKs. Our findings imply that the two classes have different evolution modes and mass
assembly histories.

\normalem
\begin{acknowledgements}
This work is based on observations taken
by the CANDELS Multi-Cycle Treasury Program with the
NASA/ESA HST, which is operated by the Association of
Universities for Research in Astronomy, Inc., under the NASA contract NAS5-26555.
This work is supported by the National Natural Science Foundation of China (NSFC, Nos.
11303002, 11225315, 1320101002, 11433005, and 11421303), the Specialized Research Fund for the
Doctoral Program of Higher Education (SRFDP, No. 20123402110037), the Strategic Priority
Research Program ``The Emergence of Cosmological Structures" of the Chinese Academy of
Sciences (No. XDB09000000), the Chinese National 973 Fundamental Science Programs
(973 program) (2015CB857004), the Yunnan Applied Basic Research Projects (2014FB155)
and the Open Research Program of Key Laboratory for Research in Galaxies and Cosmology, CAS.
\end{acknowledgements}

\label{lastpage}
\end{document}